\documentclass[aps,prb,twocolumn,a4paper,superscriptaddress,floatfix,showpacs,preprintnumbers,amsmath,amssymb]{revtex4}
\usepackage{url}
\usepackage[english]{babel}
\usepackage[utf8]{inputenc}
\usepackage{amsmath}
\usepackage{graphicx,epstopdf}
\usepackage{amssymb,epsfig,color,textcase}
\usepackage{hyperref}
\usepackage{doi}
\providecommand{\U}[1]{\protect\rule{.1in}{.1in}}

\def\feo2{FeO$_2$}
\def\feoh2{FeO$_2$H}
\def\feo2{FeO$_2$}
\def\fe2o{Fe$_2$O}
\def\etal{\textit{et al.}}

\begin{document}

\preprint{APS/123-QED}

\title{Importance of the many-body effects for structural properties of the novel iron oxide: Fe$_2$O }

\author{Alexey O. Shorikov}
\email{shorikov@imp.uran.ru}
\affiliation{M.N. Miheev Institute of Metal Physics of Ural Branch of Russian Academy of Sciences - 620990 Yekaterinburg, Russia}
\affiliation{Department of theoretical physics and applied mathematics, Ural Federal University, Mira St. 19, 620002 Yekaterinburg, Russia}
\affiliation{Skolkovo Institute of Science and Technology, 3 Nobel St., Moscow, 143026, Russia}

\author{Sergey V. Streltsov}%
\affiliation{M.N. Miheev Institute of Metal Physics of Ural Branch of Russian Academy of Sciences - 620990 Yekaterinburg, Russia}
\affiliation{Department of theoretical physics and applied mathematics, Ural Federal University, Mira St. 19, 620002 Yekaterinburg, Russia}

\date{\today}

\begin{abstract}
The importance of many-body effects on electronic and magnetic properties and stability of different structural phases was studied in novel iron oxide - Fe$_2$O. It was found that while Hubbard repulsion hardly affects the electronic spectrum of this material ($m^*/m \sim 1.2$), but it strongly changes its phase diagram shifting critical pressures of structural transitions to much lower values. Moreover, one of the previously obtained in the density functional theory (DFT) structures (P$\bar 3$m1) becomes energetically unstable if many-body effects are taken into consideration. It is shown that this is an account of magnetic moment fluctuations in the DFT+DMFT approach, which strongly contributes to modification of the phase diagram of Fe$_2$O.
\end{abstract}


\maketitle

{\bf Introduction.}~
Substantial progress in studying and prediction of novel structures under high pressure including ones of iron oxides was achieved since methods which combine {\it ab initio} calculations with algorithms of crystal structure generation were developed. 
In their seminal work Weerasinghe and co-authors obtained the set of different structures for various iron oxides, which can be reliazed at high pressures~\cite{Weerasinghe2015}. Some of these materials, e.g. FeO$_2$ and Fe$_2$O do not exist at ambient conditions. Further experimental investigations did found evidence of FeO$_2$ formation at pressures $\sim 76$ GPa and very high temperature $\sim 1800$ K exactly in previously predicted crystal structure~\cite{Hu2016}. This discovery opened up the Pandora's box, since stabilization of novel iron oxides, which physical properties can be very different from conventional ones is very important for the Geoscience. In effect, this result even led to reexamination of different geological models describing our planet, reconsideration of the water role and composition of the Earth's core-mantle  boundary~\cite{Hu2016,Yagi2016,Liu2017,Thompson2017,Mao2017,Pushcharovsky2019}.
This impressive progress induced by the first principle calculations, however, raised many questions as those related to account of other ingredients, which exist in the Earth's interior (such as e.g hydrogen~\cite{Nishi2017,Hu2017,Shorikov2018,Liu2019,Shorikov2020,Lu2018,Zhang2018}), and others connected with applicability of the density functional theory (DFT) calculations for predictions of high-pressure phases and analysis of their physical properties. In particular it is known from long ago that strong electron Coulomb correlations, may dramatically change not only structural stability~\cite{Shorikov2018b}, but also electronic and magnetic properties of iron oxides and sulphides~\cite{Streltsov2017b,Ushakov2017,Shorikov2015a,Skorikov2015}.
Recently importance of  weak Coulomb correlations for description of phase diagram was demonstrated even for elemental Ca~\cite{Novoselov2020}.

In the present paper we studied this effect on example of Fe$_2$O, which was predicted by DFT to be more stable than assemblage of conventional FeO and pure iron at pressures larger than 270 GPa~\cite{Huang2018} using state-of-the art DFT+DMFT method, which was previously used for description of such materials as Fe~\cite{Leonov}, FeO$_2$~\cite{Streltsov2017b}, FeS~\cite{Ushakov2017} and many others under pressure.  We show that while there is only a relatively weak effective mass renormalization ($m^*/m \approx 1.25$) and spectral and magnetic properties seems to be describable by the DFT, nevertheless many-body effects drastically affect phase diagram of Fe$_2$O. In particular the P$\bar{3}$m1 phase predicted by DFT was found unstable in the DFT+DMFT calculations. More important we found that the I4/mmm phase should be stable at much lower pressures $\approx$ 170 GPa and transforms into assemblage of Pnma FeO and hcp-Fe with decrease of temperature.

\begin{figure}[b!]
  \centering
  \includegraphics[width=0.480\textwidth]{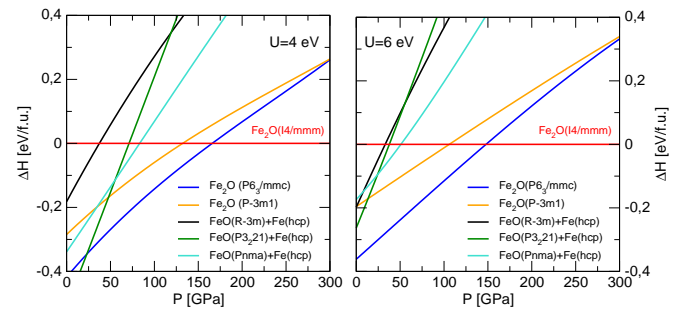}
  \caption {(Color online) Plots of relative enthalpy against pressure for Fe$_2$O and assemblage of FeO and hcp-Fe as obtained in DFT+DMFT at 800~K calculations for $U$=4 eV (left panel) and $U=6$ eV (right panel). The enthalpy of I4/mmm phase is shown as the zero line.}
  \label{ent-DMFT-u4-800}
\end{figure} 

{\bf Computational details.}~  
The DFT calculations were carried out with the pseudopotential VASP package~\cite{Kresse1996} using the PBE exchange-correlation functional~\cite{Perdew1996}. At the first stage of our study we constructed series of Fe$_2$O crystal structures with three different cell symmetries, namely: I4/mmm, P$6_3$/mmc and P$\bar{3}$m1 and different volumes from $\approx$ 8 to 14 $\AA^3$ per iron atom, FeO in R$\bar 3$m and Pnma and pure iron in hcp structures using full structural relaxation. Then the Wannier function projection procedure~\cite{Korotin2008} was used to extract the non-interacting DFT Hamiltonian $H_{DFT}$. Specific crystal structures of Fe$_2$O results in complicated picture of band structure where both $s-$ and $d-$states of iron are presented in the vicinity the Fermi level. Therefore minimal set of the Wannier functions, which has to be included to reproduce DFT bands is $s$- and $d$-states of Fe and $p$-states of O.

\begin{figure}[t!]
  \centering
  \includegraphics[width=0.480\textwidth]{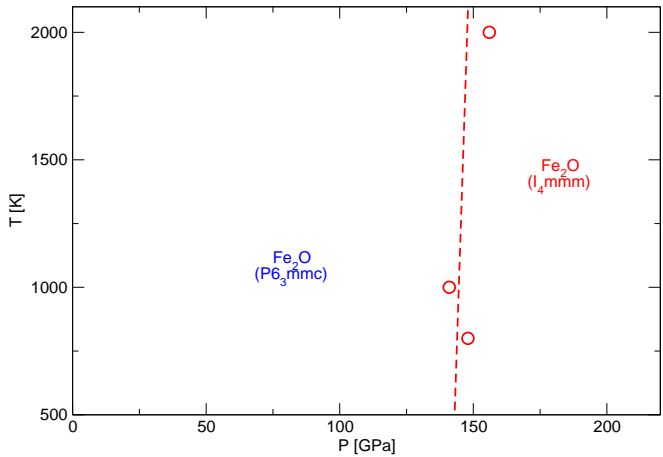}
  \caption {(Color online) Phase diagram obtained in DFT+DMFT for $U$= 6 eV. Circles shows the crossing points of P$6_3$/mmc and I4/mmm enthalpies at different temperatures. Dashed line is shown as a guide for eyes. }
  \label{PD-DMFT}
\end{figure}

 Dynamic nature of Coulomb correlations were taken into account in frameworks of  the DFT+DMFT method\cite{Anisimov1997}.  This approach was successfully used in investigating different magnetic phenomena, including spin state transitions under high pressure~\cite{Dyachenko2012,Skorikov2015} and many others. In contrast to DFT+U method, DFT+DMFT provides more accurate treatment of Coulomb interaction considering frequency dependence of the self-energy and simulating a paramagnetic state which is expected at such high temperatures and pressures. The full many-body Hamiltonian to be solved in DFT+DMFT has the form:
\begin{equation}
\hat H= \hat H_{\mathrm{DFT}}- \hat H_{\mathrm{DC}}+\frac{1}{2}\sum_{i,m,m',\sigma,\sigma^{\prime}}
U^{\sigma,\sigma^{\prime}}_{m,m'}\hat n_{i,m,\sigma,}\hat n_{i,m',\sigma^{\prime}}.
\label{eq:ham}
\end{equation}
Here $U^{\sigma,\sigma^{\prime}}_{m,m'}$ is the Coulomb interaction matrix and $\hat n_{im,\sigma}$ is the occupation number operator for the $d$ electrons with orbital and spin indices $m,\sigma$ on the $i$-th site. The elements of $U_{m,m'}^{\sigma\sigma'}$ matrix are parameterized by the on-site Hubbard parameter $U$ and Hund's intra-atomic exchange $J_H$ according to the procedure described in Ref~\cite{Lichtenstein1998}.  $U$ parameter stands for the energy cost of double occupancy of a local orbital. Its value changes the orbital localization and therefore affects on the total energy and phase stability. The value of $U$ parameter for iron oxides used in previous works varies from 4.6 eV~\cite{Pickett1998} to 8 eV~\cite{Leonov2020}. In our calculations we set $U=6$~eV and $J_H=0.89$~eV. These values were successfully used previously to describe electronic properties of various Fe oxides~\cite{Ushakov2017,Anisimov2008,Streltsov2017b}. However, it was shown that $U$ should decrease under pressure as a result of more efficient screening due to increase of hybridization strength and decrease of localization~\cite{Skorikov2015, Dyachenko2016}.  For testing the dependence of phase stability on $U$ we also performed additional calculations for $U=4$ eV, their results are shown in Fig.~\ref{ent-DMFT-u4-800} (left panel) and in supplementary materials. All results, if not stated otherwise, presented below were obtained for $U=6$ eV. 

The term $\hat H_{\mathrm{DC}}$ in \eqref{eq:ham} stands for the so called double-counting correction i.e. the $d$-$d$ interaction energy already accounted by DFT. We choose the double-counting term in the form $\hat H_{\mathrm{DC}}=\bar{U}(N_{d}-\frac{1}{2})\hat{I}$~\cite{Anisimov1997}. Here $N_{d}$ is the total self-consistent number of $d$ electrons obtained within the DFT+DMFT, $\bar{U}$ is the average Coulomb parameter for the $d$ shell and $\hat I$ is the identity operator.
\begin{figure}[t!]
  \centering
  \includegraphics[width=0.450\textwidth]{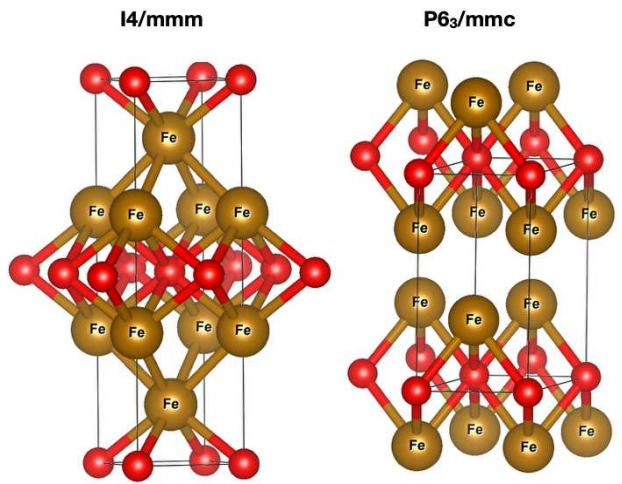}  
  \caption {(Color online) Crystal structures for Fe$_2$O as obtained after relaxation in the DFT calculations.  Fe and O atoms are shown by brown and blue balls, respectively.}
  \label{fig:str}
\end{figure}

The effective DMFT impurity problem was solved by the hybridization expansion Continuous-Time Quantum Monte-Carlo method (CT-QMC-HYB)~\cite{Gull2011} as implemented in the AMULET simulation package~\cite{AMULET}. We apply diagonal solver with the density-density form of Coulomb correlation term. In the case of I4/mmm crystallographic axes coincide with direction to next nearest neighbours and no additional transformation were needed.  All DFT Hamiltonian for P$6_3$/mmc and $P\bar{3}m1$ cells were additionally diagonalized which correspond to transformation into trigonal coordinate system in which Hamiltonians become diagonal. For the sake of simplicity we used notation of orbitals from trigonal symmetry: $e_g^{\pi}$, $a_{1g}$ and $e_g^{\sigma}$.  The calculations were performed for the paramagnetic state at a series of electronic temperatures ($T=1/\beta$) from 589~K to 2000~K. Total energy was calculated within the DFT+DMFT as described in Ref.~\cite{Amadon2006}. Next, the DFT and DFT+DMFT total energies were fitted using third-order Birch-Murnaghan  equation of states~\cite{Birch1947} for calculating enthalpies and constructing the P-T phase diagram.  To compute the spectral properties and renormalizations of the quasiparticle mass we used the real-axis self-energy $\Sigma(\omega)$ obtained by the Pad\'e analytical continuation  procedure~\cite{pade}.

{\bf Structural stability of different phases of Fe$_2$O.}~Different structural phases of Fe$_2$O were studied by Weerasinghe et al., and Huang et al. using DFT calculations~\cite{Weerasinghe2015,Huang2018}. According to Ref.~\cite{Weerasinghe2015} Fe$_2$O is formed at pressures $\sim$ 100 GPa. They found that below 180 GPa~\cite{Weerasinghe2015} (212 GPa~\cite{Huang2018}) it has the P$6_3$/mmc structure. Then it transforms to the P$\bar 3$m1 phase and finally at 288 GPa~\cite{Weerasinghe2015} (266 GPa~\cite{Huang2018}) to the I4/mmm structure. It was shown in Ref.~\cite{Huang2018} that above 266 GPa I4/mmm phase of Fe$_2$O is not only statically and dynamically stable, but also becomes more energetically favourable than the assemblage of hcp-Fe and R$\bar 3$m-FeO. Thus, one might expect that this is Fe$_2$O with I4/mmm symmetry that realizes at very high pressure. Our DFT results basically agree with this picture with transition pressures $\sim$150 and $\sim$250 GPa~\cite{SI}.

However, accounting for many-body effects in the DFT+DMFT calculation scheme substantially modifies phase diagram of Fe$_2$O. First of all, one readily sees from Fig.~\ref{ent-DMFT-u4-800}, where relative enthalpy of different phases at $T=800$ K is plotted, that they decrease critical pressure for transition to the I4/mmm structure on $\sim 100$ GPa, down to 150 GPa (on $80$ GPa, down to 170 GPa for $U$= 4 eV). Moreover, many-body effects strongly destabilize P$\bar 3$m1 phase so that it does not realize at investigated temperature and pressure range for any reasonable $U$ value.
\begin{figure}[t!]
  \centering
  \includegraphics[width=0.450\textwidth]{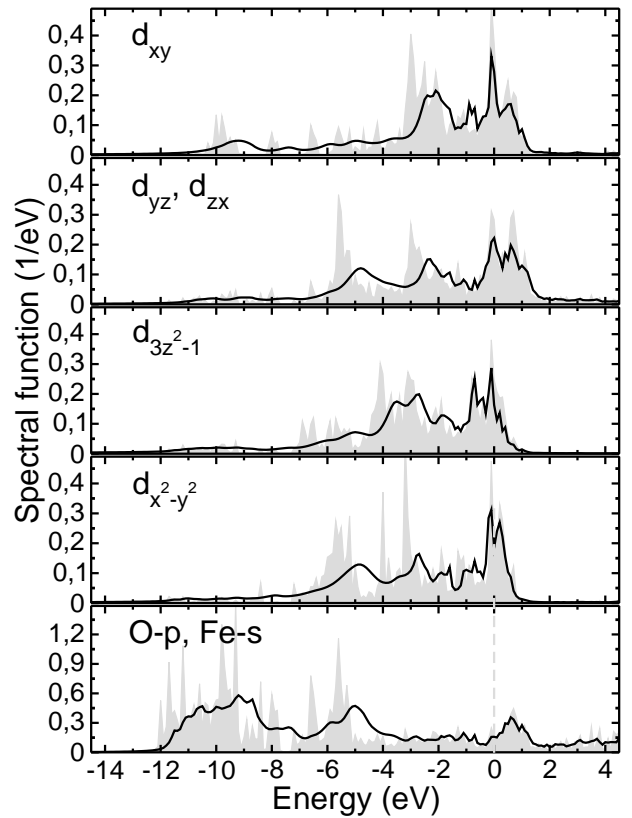}
  \caption {Orbitally resolved spectral function for the I4/mmm structure as obtained in the DFT+DMFT for $P=172$ GPa and at $T=580$~K (solid lines). DFT DOSes are shows for comparison as shaded areas. Fermi energy is set to zero (shown as grey dashed line).}
  \label{fig:dos:I4mmm}
\end{figure}

In addition to calculations of relative stability of different structures we also computed enthalphy of Fe$_2$O formation taking similarly to Ref.~\cite{Huang2018} assemblage of FeO and hcp-Fe as a reference point. Three structural phases of FeO (R$\bar 3$m, P3$_2$21, and Pnma) were taken into account according to Ref.~\cite{Weerasinghe2015}. Corresponding enthalpies are presented again in Fig.~\ref{ent-DMFT-u4-800}.  One might see that many-body effects also strongly stabilize the I4/mmm structure. E.g. according to DFT+DMFT calculations at temperatures of $\sim$ 600 K Fe$_2$O can be formed only for pressures larger $\sim$ 150 GPa, while this critical pressure was much higher in DFT: $\sim 250$ GPa (see Fig. S1 in Ref.~\cite{SI}).  As it has been mentioned above this critical pressure shifts to higher values for smaller $U$. One can also notice that for $U=4$ eV Fe$_2$O decomposes onto FeO and elemental Fe at slightly smaller pressure.

Full phase diagram is presented in Fig.~\ref{PD-DMFT}. We see that, below $\sim 150$ GPa P$6_3$/mmc phase of Fe$_2$O turns out to be statically stable at temperatures higher than 800~K, while above this pressure Fe$_2$O has the I4/mmm structure. Note that for lower $U$ value at lower pressure and low temperature Fe$_2$O decomposes into  mixture of FeO in P3$_2$21 structure and pure iron. Since this phase transition is sensitive to parameter choice more accurate estimation of Coulomb parameters is needed. Note also that this transition occurs at rather low temperatures, which are seemingly irrelevant for the conditions of outer Earth's core.

{\bf Crystal structure of stable phases}.  The crystal structure of the I4/mmm phase is shown in left panel of Fig.~\ref{fig:str}. Basically it can be described as two Fe square lattices sandwiched by O square lattices. These two Fe lattices are shifted with respect to each other in such a way that Fe ions of one lattice are on the top of voids of another one. It is worth mentioning that in fact distance between Fe ions within the square lattice is even larger than Fe-Fe bond distance between planes, e.g. for $P=350$ GPa the first one is 2.137 \AA, while the second is 2.063 \AA. Thus these are not two isolated planes, but a single object. It also has to be mentioned that both distances are smaller than Fe-Fe distance in metallic Fe at ambient conditions (2.482 \AA~\cite{Streltsov-UFN}). These Fe-Fe distance are still larger than Fe-O distance (1.88 \AA), but the difference is not that large. 
 \begin{figure}[t!]
  \centering
  \includegraphics[width=0.450\textwidth]{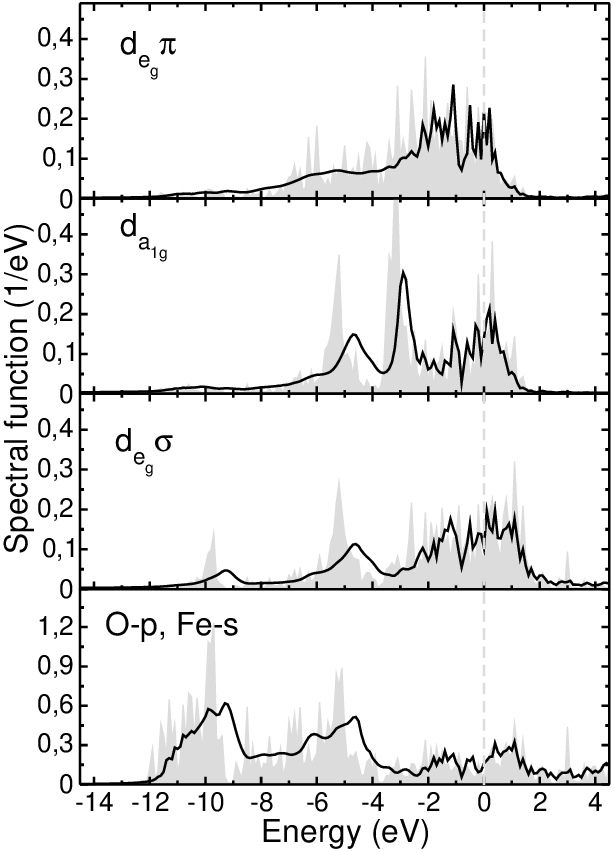}
   \caption {Orbitally resolved spectral function for the P$6_3$/mmc structure as obtained in the DFT+DMFT for $P=173$ GPa and at $T=580$~K (solid lines). DFT DOSes are shows for comparison as shaded areas. Fermi energy is set to zero (shown as grey dashed line).}
  \label{fig:dos:P63mmc}
\end{figure}

Crystal structure of the P$6_3$/mmc phase reminds I4/mmm, see Fig.~\ref{fig:str}. There are again two planes of Fe ions, but now they form triangular (not square) lattice with Fe-Fe distance 2.352 \AA~and interplane Fe-Fe distance 2.252 \AA~at $P=100$ GPa. Each Fe ion stands on the top or bottom of three (four in I4/mmm) oxygens. The Fe-O distances at $P=100$ GPa are 1.908 \AA.

{\bf Electronic and magnetic properties.}~ While we have clearly seen that many-body effects strongly affect phase stability in Fe$_2$O it is interesting how they change electronic and magnetic properties.   

Fig.~\ref{fig:dos:I4mmm} shows orbitally resolved spectral function for the I4/mmm structure at 170 GPa as obtained in DFT+DMFT compared with densities of states (DOS) calculated in DFT.  The Fe $3d$ bandwidth exceeds 8 eV at the DFT level. The DFT+DMFT calculations demonstrate that correlations have only a minor effect on spectral spectral properties of such broad bands. Spectral function obtained in DFT+DMFT at 580~K is rather similar to the DFT density of states, but is slightly smoothed by temperature. Effect of finite temperature taken into account in DMFT leads to broadening of all bands and to smearing of narrow peaks in the spectrum. The shape of spectral function corresponding to Fe $3d$ states undergoes the most dramatic changes. The peak at the Fermi level is smoothed and the width of whole the Fe $3d$ band slightly decreases due to correlation effects. Table~\ref{Table1} shows the calculated effective mass renormalization $m^*/m=Z^{-1}$, where $Z=(1-\frac{d\Sigma(\omega)}{d\omega}|_{\omega=0})^{-1}$. The obtained average value of effective mass enhancement $m^*/m$ is rather small ($\sim$1.2) which indicates weak or intermediate correlation strength regime.  
One can see that Fe$_2$O remains metallic and there is a small narrowing of spectral function, which is in agreement with $m^*/m$ values presented below.

\begin{table}[b]
\caption
{
Orbitally-resolved enhancement of the electron mass $m^{*}/m$ 
in \fe2o for different orbitals of the $d$ shell as obtained by DFT+DMFT at $T=580$~K and $P\approx171$~GPa.
}
\centering
\begin{tabular}{cccccc}
\hline
\hline
            Phase&   $d_{xy}$    &  $d_{yz}$,$d_{zx}$  &   $d_{3z^2-1}$ &  $d_{x^2-y^2}$& $ \langle {m^{*}}/{m}\rangle $  \\
\hline
I4/mmm        & $1.20$ & $1.18$ & $1.15$ & $1.16$ & $1.18$ \\
P$\bar{3}$m1   & $1.15$ & $1.15$ & $1.18$ & $1.17$ & $1.16$ \\

P$6_3$/mmc      & $1.16$ & $1.16$ & $1.19$ & $1.18$ & $1.17$ \\
\hline
\hline
\end{tabular}
\label{Table1}
\end{table}

Electronic structure of the P$6_3$/mmc phase is rather similar to what we had for the I4/mmm structure. The width ($W$) of $3d$ bands is of order of 8-10 eV. Correlation effects are not pronounced, $m^*/m \sim 1.16$, and again result only in thermal broadening, see Fig.~\ref{fig:dos:P63mmc}. It also has to be mentioned that the correlation strength in Fe$_2$O and in FeO at pressures $\sim 200$ GPa is of the same order, in FeO $m^*/m \sim 1.3$. 

It has to be mentioned that the difference between DFT and DFT+DMFT is not only in account of strong Hubbard correlations and the fact that effective mass renormalization is small (correlations are weak) does not contradict to a strong modification of the phase diagram in DMFT discussed previously. In addition to explicit account of on-site electronic correlation effects DMFT calculations also include various thermal and quantum fluctuations in consideration. E.g. DMFT can treat a true paramagnetic ground state with short-range spin correlations, but without any long-range magnetic order. This is very different from DFT, which simulates such states by a nonmagnetic (zero magnetic moments) state or by some ferro- or antiferromagnetic state (we do not consider here some special DFT calculations with large supercell). We will show further on that the magnetic moments do exist in Fe$_2$O, they strongly fluctuate, but this affects the physical properties of the material under consideration.
\begin{figure}[hb!]
  \centering
     \includegraphics[width=0.450\textwidth]{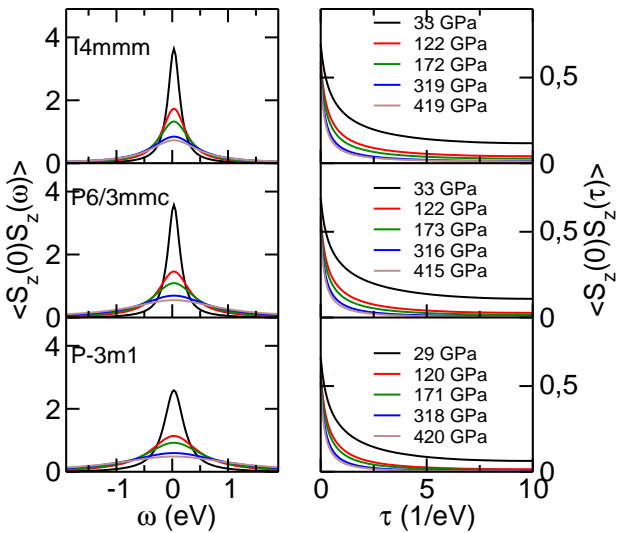}
  \caption {(Color online) Local spin-spin correlation function $\langle S_z(0) S_z(\tau)\rangle$ from 0 to $\beta/2$ (right) obtained within DFT+DMFT method for Fe$_2$O for T=580~K. Left panels show $\langle S_z(0) S_z(\omega)\rangle$ on real energies.}
  \label{fig:corr}
\end{figure}

The local spin-spin correlation function $\langle S_z(0)S_z(\tau)\rangle$ on the imaginary time ($\tau$) axis characterizes the lifetime of the local moment: if the magnetic moments are localized, this correlation function is constant: $\approx S^2$, on the other hand, the imaginary time dependence of this correlation function indicates the delocalization of spin moments. The behavior of $\langle S_z(0) S_z(\tau)\rangle$ for three phases of Fe$_2$O is shown in Fig.~\ref{fig:corr} (right panels). Rapid drop of this correlation function with $\tau$ implies that magnetic moments are almost itinerant in all phases and moreover localization degree decreases with pressure which agrees well with low $m^*/m$ value and this is exactly what one expects to have at high pressure due to bands broadening. The observed behavior of these correlation functions is traced in analytical continuation of them on real energy axis $\langle S_z(0) S_z(\omega)\rangle$, see left panel of Fig.~\ref{fig:corr}. The half-width of this function is inverse proportional to lifetime of local magnetic moment~\cite{Igoshev2013}. The most interesting result is that P$\bar3$m1 demonstrates substantially lower degree of localization than the I4/mmm and P$6_3$/mmc phases at similar pressures. As we will see below this substantially decreases the total energy of this phase.
\begin{figure}[hb!]
  \centering
  \includegraphics[width=0.450\textwidth]{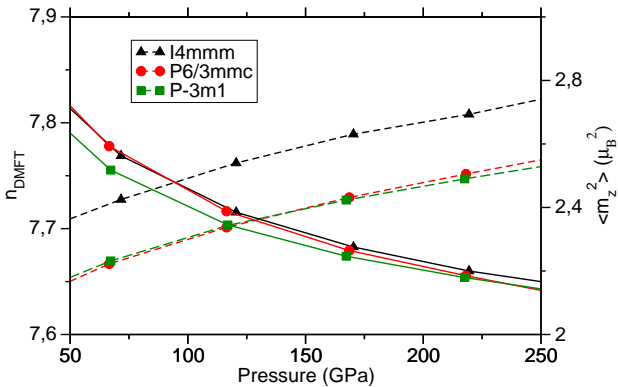}
  \caption {(Color online) Instant squared magnetic moments (right axis) for I4/mmm (black triangles) P$6_3$/mmc (red circles) and P$\bar3$m1 (green squares) as obtained in DFT+DMFT at $T=1160$~K as a function of pressure are shown with solid lines. Dashed lines shows corresponding total number of electrons on $d$-shell (left axis).}
  \label{fig:mag_mom}
\end{figure}

In all studied structures Fe ions were found to adopt the low spin (LS) state and the ground state is paramagnetic (no long range magnetic ordering). Fig.~\ref{fig:mag_mom} shows pressure dependence of  the instant squared magnetic moments calculated within DFT+DMFT and the total number of $d$-electrons at $T=1160$~K for all phases. The magnitude of average local moment $\langle m_z^2\rangle$ decreases gradually with pressure, e.g. in the P$6_3$/mmc phase it is $\sim$ 2.9 $\mu_B^2$ at ambient pressure (AP) and 2.0 $\mu_B^2$ at $\sim$ 400~GPa. Similar behavior was observed for the I4/mmm phase (2.8 $\mu_B^2$ and 2.0 $\mu_B^2$). 
The total number of $d$-electrons increases since the band width growths with pressure as a consequence of the increase of orbital overlap implying decrease of $U/W$ ratio. Here again we see that P$\bar3$m1 demonstrates lower magnitude of local moment than I4/mmm and  P6$_3$/mmc which show slightly higher and almost equal value of $\langle m_z^2 \rangle$. 

This behavior correlates with the fact that the P$\bar3$m1 phase turns out to be more itinerant at the pressure up to $\approx$ 220 GPa than other phases as it was explained above. This phase could have been realized in pressure range from 150 to $\sim$260 GPa in DFT, but smaller instant magnetic moment strongly destabilizes it in DMFT.  Indeed for itinerant magnets contribution of the magnetic energy is known to be proportional to $-Im^2/4$, where $m$ is the magnetic moment and $I$ is the Stoner parameter, which can be roughly approximated by Hund's intra-atomic exchange, i.e. $\sim$ 1 eV. Taking realistic estimates of $\langle m_z^2 \rangle$ from the DFT+DMFT calculations we see that difference in magnetic energy between the P$\bar3$m1 and other phases is $\sim$ 0.12 eV/atom. This contribution strongly increases enthalpy of the P$\bar3$m1 phase and therefore explains its destabilization in DMFT calculations. While I4/mmm has the largest instant magnetic moment in pressure range from 150 to $\sim$260 GPa we can not exclude that other factors such as e.g. some specific features of the electronic structure may also affect the resultant energetics.

{\bf Conclusions.}~
We demonstrated that accounting for many-body effects is crucial for correct description of phase stability of Fe$_2$O at high pressures corresponding to the Earth's outer core. The DFT+DMFT calculations show that magnetic moments in this material are nearly itinerant and electronic mass enhancement is small $m^*/m \approx 1.2$ which implies weak or intermediate correlation strength regime. Spectral function remains almost unchanged with respect to DFT, but smoothed and narrowed in vicinity of the Fermi level. On the other hand many-effects have strong impact on the phase diagram. The stability field of the I4/mmm phase is shifted towards lower pressures on $\approx$ 80 GPa (with respect to DFT). The P$\bar3$m1 phase becomes unstable in the whole pressure range mostly due to lower magnitude of magnetic moment and hence lower impact of magnetic energy. 

{\bf Acknowledgments.}~
 The  DFT+DMFT calculations were supported by the Russian Science Foundation (Project No. 19-72-30043). The calculations of phase diagram were performed within project ``Quantum'' No. AAAA-A18-118020190095-4 and contract 02.A03.21.0006 of the Russian Ministry of Science and High Education.

\newpage
{\bf Supplemental information}
\onecolumngrid

\begin{figure}[hb!]
  \centering
  \includegraphics[width=0.90\textwidth]{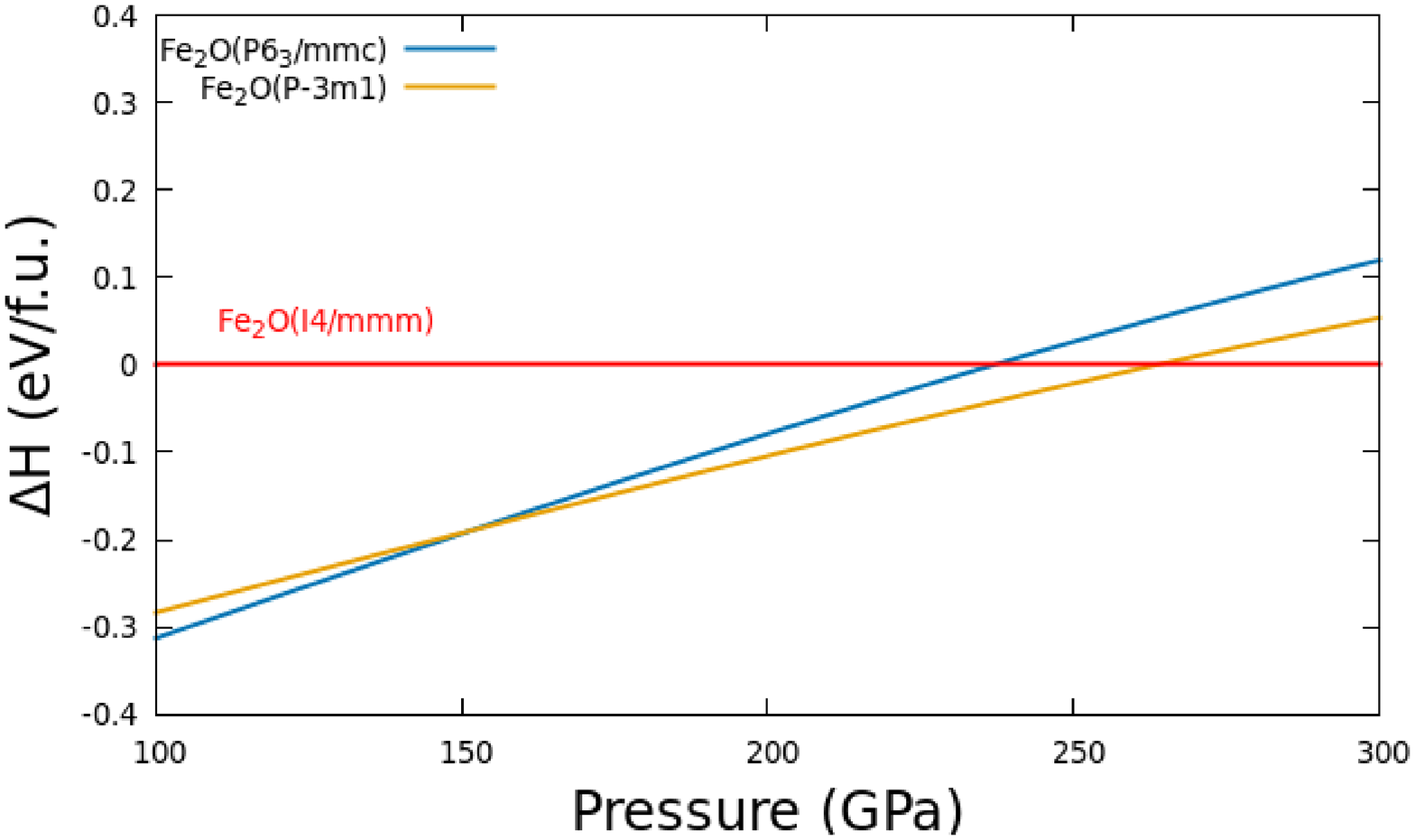}
  \caption {(Color online) Plots of relative enthalpy against pressure for Fe$_2$O as obtained in DFT. The enthalpy of I4/mmm phase is shown as the zero line.} 
  \label{fig:PD-DFT}
\end{figure}

\begin{figure}[b!]
  \centering
  \includegraphics[width=0.90\textwidth]{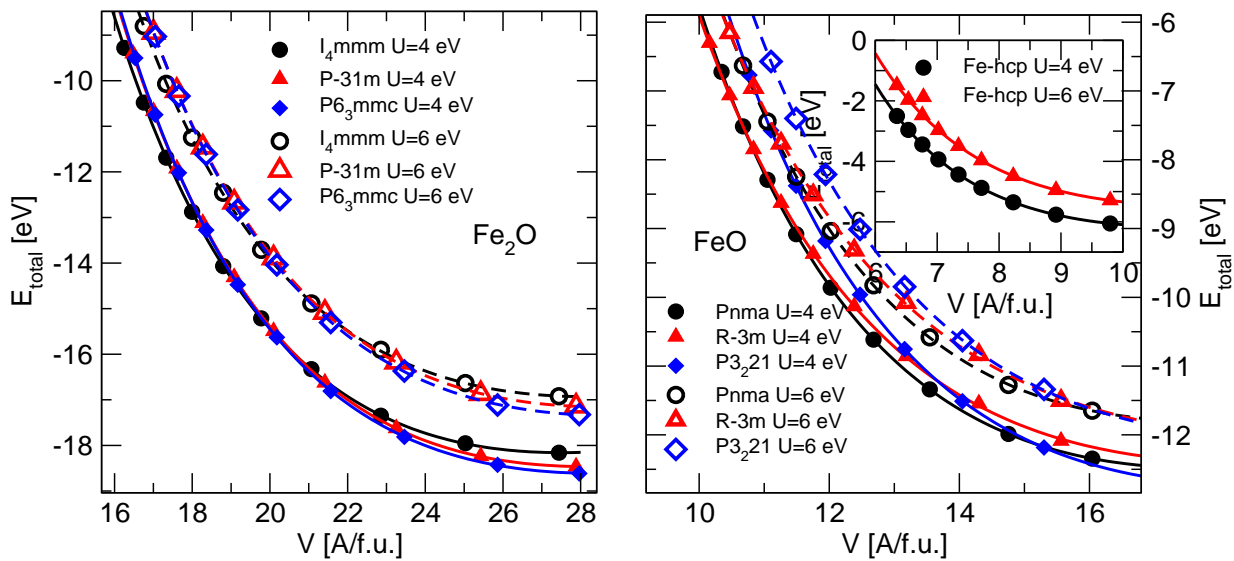}
  \caption {(Color online) Plots of DFT+DMFT total energy at T=800 K  against cell volume for Fe$_2$O (left panel) and  FeO (right panel).  Inset shows total energy for hcp-Fe.  Solid and dashed lines shows the result of fitting with Birch-Murnaghan EoS for $U$=4 and 6 eV, respectively.}
  \label{etot-800}
\end{figure}

\begin{figure}[hb!]
  \centering
  \includegraphics[width=0.90\textwidth]{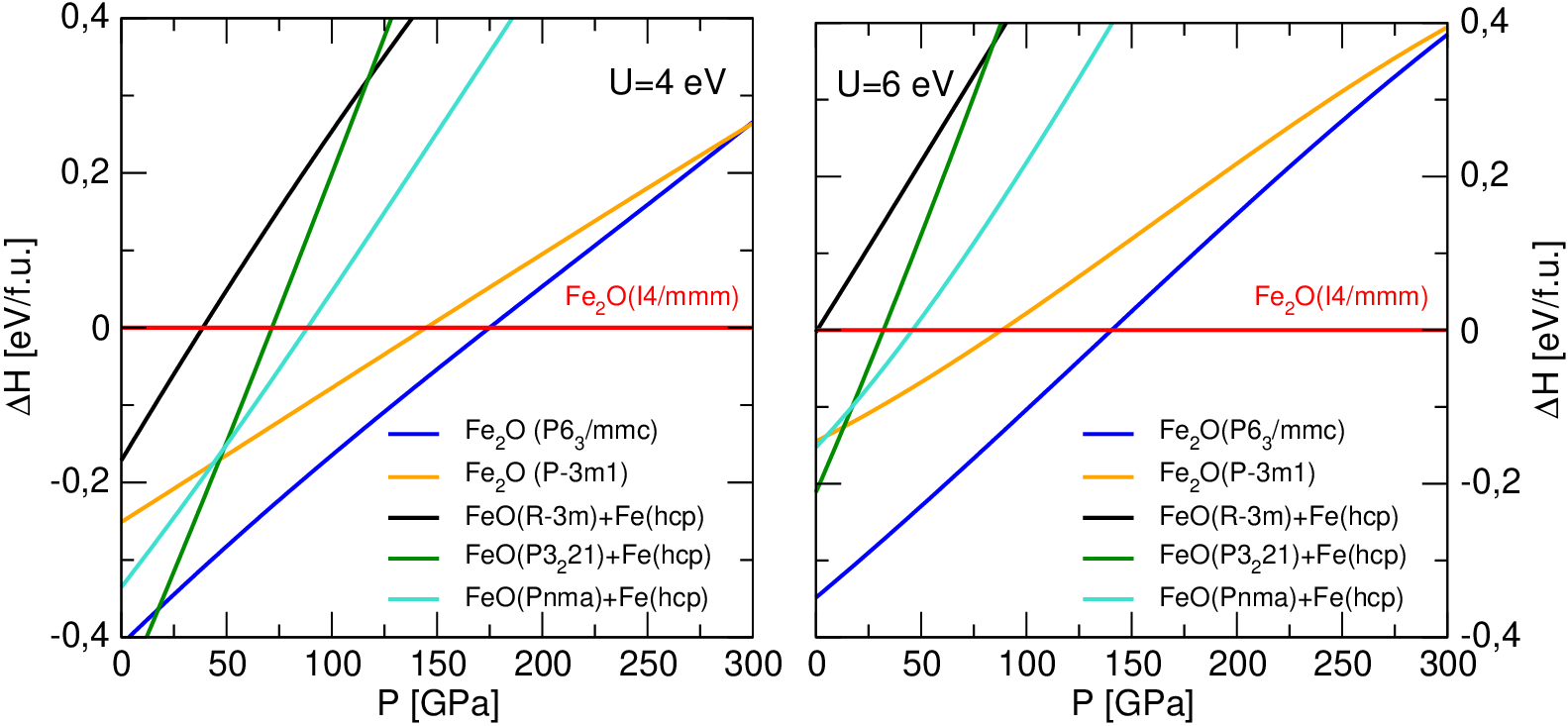}
  \caption {(Color online) Plots of relative enthalpy against pressure for Fe$_2$O and assemblage of FeO and hcp-Fe as obtained in DFT+DMFT at 1000~K calculations for U=4 eV (left panel) and U=6 eV (right panel). The enthalpy of I4/mmm phase is shown as the zero line.}. 
  \label{fig:PD-DFT+DMFT_b10}
\end{figure}

\begin{figure}[hb!]
  \centering
  \includegraphics[width=0.90\textwidth]{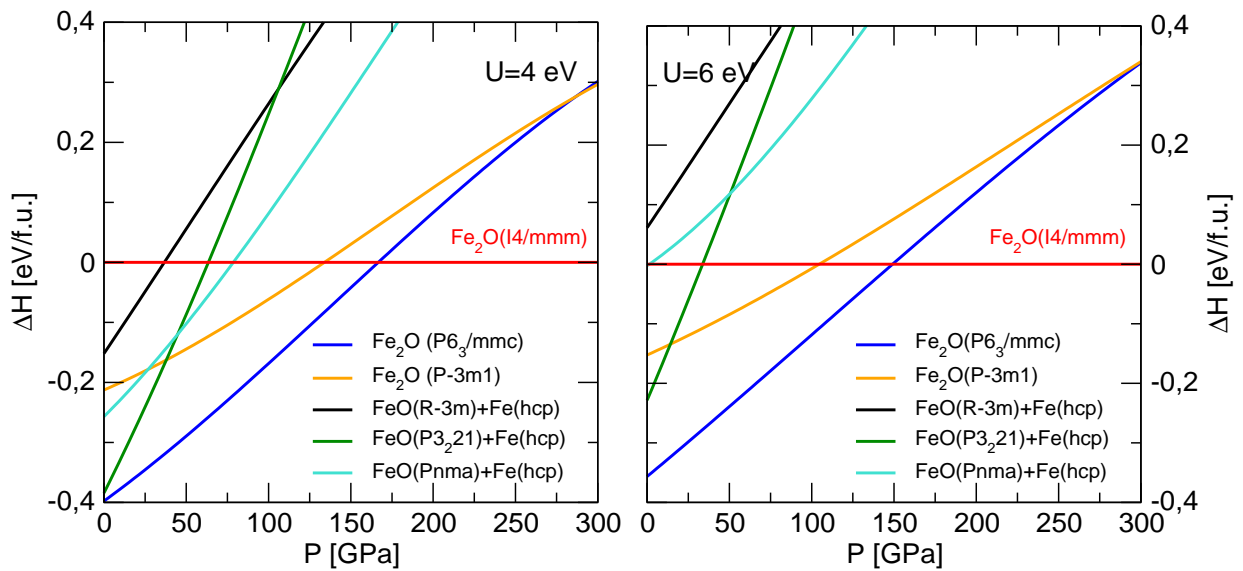}
  \caption {(Color online) Plots of relative enthalpy against pressure for Fe$_2$O and assemblage of FeO and hcp-Fe as obtained in DFT+DMFT at 2000~K calculations for U=4 eV (left panel) and U=6 eV (right panel. The enthalpy of I4/mmm phase is shown as the zero line.}. 
  \label{fig:PD-DFT+DMFT_b5}
\end{figure}

\begin{figure}[t!]
  \centering
  \includegraphics[width=0.480\textwidth]{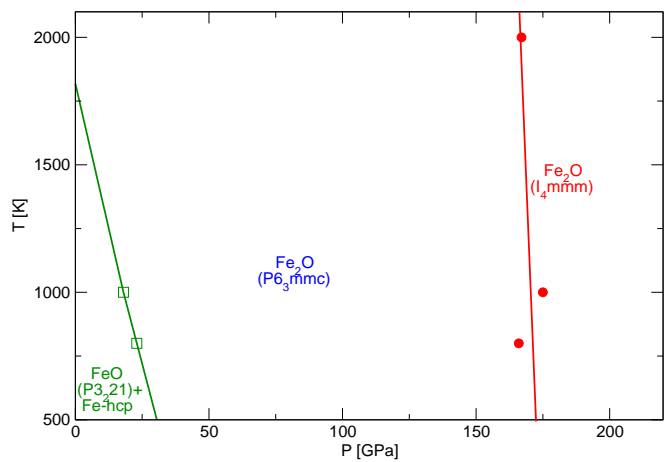}
  \caption {(Color online) Phase diagram obtained in DFT+DMFT for $U$= 4 eV.}
  \label{PD-DMFT}
\end{figure}


\begin{thebibliography}{99}

\bibitem{Weerasinghe2015} G. L. Weerasinghe, C. J. Pickard, and R. J. Needs, J. Phys. Condens. Matter \textbf {27}, 455501 (2015).

\bibitem{Hu2016} Q. Hu, D. Y. Kim, W. Yang, L. Yang, Y. Meng, L. Zhang, and H. Mao, Nature \textbf {534}, 241 (2016).

\bibitem{Yagi2016} T. Yagi, Nature \textbf {534}, 183 (2016).

\bibitem{Liu2017} J. Liu, Q. Hu, D. Y. Kim, Z. Wu, W. Wang, Y. Xiao, P. Chow, Y. Meng, V. B. Prakapenka, H. Mao, and W. L. Mao, Nature \textbf {551}, 494 (2017).

\bibitem{Mao2017} H. K. Mao, Q. Hu, L. Yang, J. Liu, D. Y. Kim, Y. Meng, L. Zhang, V. B. Prakapenka, W. Yang, and W. L. Mao, Natl. Sci. Rev.\textbf { 4}, 870 (2017).

\bibitem{Thompson2017} E. C. Thompson, A. J. Campbell, and J. Tsuchiya, J. Geophys. Res. \textbf {122}, 5038 (2017).

\bibitem{Pushcharovsky2019} D. Y. Pushcharovsky, Geochemistry Int. \textbf {57}, 941 (2019).

\bibitem{Nishi2017} M. Nishi, Y. Kuwayama, J. Tsuchiya, and T. Tsuchiya, Nature \textbf {547}, 205 (2017).

\bibitem{Hu2017} Q. Hu, D. Young, J. Liu, Y. Meng, L. Yang, D. Zhang, and W. L. Mao, Proc. Natl. Acad. Sci. \textbf {114}, 1498 (2017).

\bibitem{Shorikov2018} A. O. Shorikov, A. I. Poteryaev, V. I. Anisimov, and S. V Streltsov, Phys. Rev. B \textbf {98}, 165145 (2018).

\bibitem{Liu2019} J. Liu, Q. Hu, W. Bi, L. Yang, Y. Xiao, P. Chow, Y. Meng, V. B. Prakapenka, H. Mao, and W. L. Mao, Nat. Commun. \textbf {10}, 153 (2019).

\bibitem{Shorikov2020} A. O. Shorikov, S. L. Skornyakov, V. I. Anisimov, S. V. Streltsov, and A. I. Poteryaev, Molecules \textbf {25}, 2211 (2020).

\bibitem{Lu2018} C Lu, M. Amsler, and C. Chen Phys. Rev. B \textbf {98}, 054102 (2018)

\bibitem{Zhang2018} J. Zhang, J. Lv, H. Li, X. Feng, C. Lu, S.A.T. Redfern, H. Liu, Y. Ma, Phys. Rev. Lett. \textbf {21}, 255703 (2018)
%
\bibitem{Shorikov2018b} A. O. Shorikov, V. V. Roizen, A.R. Oganov and V. I. Anisimov, Phys. Rev. B {\bf 98}, 094112 (2018)
%
\bibitem{Streltsov2017b} S. V. Streltsov, A. O. Shorikov, S. L. Skornyakov, and A. I. Poteryaev, Sci. Rep. \textbf {7}, 13005 (2017).
%
\bibitem{Anisimov2008}  V. I. Anisimov, D. M. Korotin, S. V. Streltsov, A. V. Kozhevnikov, J. Kunes, A. O. Shorikov, and M. A. Korotin, JETP Lett. {\bf 88}, 729 (2008).
%
\bibitem{Ushakov2017} A.V. Ushakov,  A.O. Shorikov, V.I. Anisimov, N.V. Baranov,  S.V. Streltsov,  Phys. Rev. B {\bf 95}, 205116 (2017).
%
\bibitem{Shorikov2015a} A. O. Shorikov, A. V. Lukoyanov, V. I. Anisimov, and S. Y. Savrasov, Phys. Rev. B {\bf 92}, 035125 (2015).
%
\bibitem{Skorikov2015} N. A. Skorikov, A. O. Shorikov, S. L. Skornyakov, M. A. Korotin, and V. I. Anisimov, J. Phys.: Condens. Matter {\bf 27}, 275501 (2015).

\bibitem{Novoselov2020}  D. Novoselov, D. Korotin, A. Shorikov, A. Oganov, V. Anisimov, J. Phys: Cond. Matt. {\bf 32} 445501 (2020).

\bibitem{Leonov} I. Leonov, A. I. Poteryaev, V. I. Anisimov, and D. Vollhardt
Phys. Rev. Lett. {\bf 106}, 106405 (2011).


\bibitem{Huang2018} S. Huang, X. Wu and S. Qin, Scientific Reports {\bf 8}, 236 (2018).

\bibitem{Kresse1996} G. Kresse and J. Furthm\"{u}ller,
    Physical Review B \textbf {54}, 11169 (1996).

\bibitem{Perdew1996} J.P. Perdew, K. Burke, and M. Ernzerhof,
    Physical Review Letters \textbf {77}, 3865 (1996).

\bibitem{Korotin2008} D. Korotin, A. V. Kozhevnikov, S. L. Skornyakov, I. Leonov, N. Binggeli, V. I. Anisimov, and G. Trimarchi, Eur. Phys. J. B \textbf {65}, 91 (2008).
%
%
\bibitem{Anisimov1997} V.I. Anisimov \etal, 
    Journal of Physics: Condensed Matter \textbf {9}, 7359 (1997).

\bibitem{Dyachenko2012} A.A. Dyachenko, A.O. Shorikov, A.V. Lukoyanov and V.I. Anisimov, JETP Letters B {\bf 96}, 56-60 (2012).
%

\bibitem{Lichtenstein1998} A. I. Lichtenstein and M. I. Katsnelson,
Phys. Rev. B \textbf {57}, 6884 (1998).

\bibitem{Pickett1998} W.E. Pickett, S.C. Erwin  and E.C. Ethridge,  
Phys. Rev. B \textbf {58}, 1201-1209 (1998).

\bibitem{Leonov2020} I. Leonov, A. O. Shorikov, V. I. Anisimov and I. A. Abrikosov,
Phys. Rev. B \textbf {101}, 245144 (2020).

\bibitem{Dyachenko2016} A.A. Dyachenko, A.O. Shorikov, A.V.  Lukoyanov and V.I. Anisimov, Phys. Rev. B \textbf {93}, 245121 (2016).
%
\bibitem{Gull2011} E. Gull, A. J. Millis, A. I. Lichtenstein,
A. N. Rubtsov, M. Troyer, and P. Werner,
Rev. Mod. Phys. \textbf {83}, 349 (2011).
%
\bibitem{AMULET} A. Poteryaev, A. Belozerov, A. Dyachenko, D. Korotin,
M. Korotin, A. Shorikov, N. Skorikov, S. Skornyakov,
and S. Streltsov, “AMULET,” http://amulet-code.org.

%
\bibitem{Amadon2006} B. Amadon, S. Biermann, A.  Georges, F. Aryasetiawan, Physical Review Letters \textbf {96}, 066402 (2006).
%
\bibitem{Birch1947} F. Birch, Physical Review \textbf {71}, 809 (1947).
%
%
\bibitem{pade} H.J. Vidberg  and J.E. Serene, J. Low Temp. Phys. {\bf 29}, 179 (1977)

\bibitem{SI} A.O. Shorikov and S.V. Streltsov Supplemental materials. 


\bibitem{Streltsov-UFN} S. V. Streltsov and D. I. Khomskii, Physics-Uspekhi \textbf {60}, 1121 (2017).

\bibitem{Igoshev2013} P. A. Igoshev, A. V. Efremov, A. I. Poteryaev, A. A. Katanin, and V. I. Anisimov, Phys. Rev. B \textbf {88}, 155120 (2013).


%


%
%



%
%


\end{thebibliography}
\end{document}